\documentclass[12pt,preprint]{aastex}

\shorttitle{NVST observations of a flux rope}
\shortauthors{Yang et al.}

\begin{document}

\title{New Vacuum Solar Telescope observations of a flux rope
\\ tracked by a filament activation}

\author{Shuhong Yang\altaffilmark{1}, Jun Zhang\altaffilmark{1},
Zhong Liu\altaffilmark{2}, and Yongyuan Xiang\altaffilmark{2}}

\altaffiltext{1}{Key Laboratory of Solar Activity, National
Astronomical Observatories, Chinese Academy of Sciences, Beijing
100012, China; shuhongyang@nao.cas.cn, zjun@nao.cas.cn}

\altaffiltext{2}{Fuxian Solar Observatory, Yunnan Astronomical
Observatory, Chinese Academy of Sciences, Kunming 650011, China;
lz@ynao.ac.cn}

\begin{abstract}

One main goal of the New Vacuum Solar Telescope (NVST) which is
located at the \emph{Fuxian Solar Observatory} is to image the Sun
at high resolution. Based on the high spatial and temporal
resolution NVST H$\alpha$ data and combined with the simultaneous
observations from the \emph{Solar Dynamics Observatory} for the
first time, we investigate a flux rope tracked by a filament
activation. The filament material is initially located at one end of
the flux rope and fills in a section of the rope, and then the
filament is activated due to magnetic field cancellation. The
activated filament rises and flows along helical threads, tracking
out the twisted flux rope structure. The length of the flux rope is
about 75 Mm, the average width of its individual threads is 1.11 Mm,
and the estimated twist is 1$\pi$. The flux rope appears as a dark
structure in H$\alpha$ images, a partial dark and partial bright
structure in 304 {\AA}, while as bright structures in 171 {\AA} and
131 {\AA} images. During this process, the overlying coronal loops
are quite steady since the filament is confined within the flux rope
and does not erupt successfully. It seems that, for the event in
this study, the filament is located and confined within the flux
rope threads, instead of being suspended in the dips of twisted
magnetic flux.

\end{abstract}

\keywords{Sun: atmosphere --- Sun: evolution --- Sun: filaments,
prominences --- Sun: magnetic fields}

\section{INTRODUCTION}

Coronal mass ejections (CMEs) are large-scale eruptive phenomena of
the Sun and release a great deal of plasma and magnetic flux into
the interplanetary space, consequently disturbing the space
environment around the Earth (Gosling 1993; Webb et al. 1994). As
identified in the white light observations, the structure of a CME
consists of three parts, i.e., a bright leading front, a dark
cavity, and a bright core (Illing \& Hundhausen 1983; Chen 2011).
The dark cavity is generally deemed to be a twisted magnetic flux
rope (Gibson et al. 2006; Riley et al. 2008). The inner bright core
is widely believed to be filament matter suspended in flux rope dips
(Guo et al. 2010; Jing et al. 2010). Filament structures are quite
conspicuous in H$\alpha$ observations (Hirayama 1985; Martin 1998;
Lin et al. 2005) and their dynamic interactions can be caused by
magnetic reconnection between the filament-carrying magnetic fields
(T{\"o}r{\"o}k et al. 2011; Jiang et al. 2013). Detailed analyses
reveal that magnetic flux ropes play a critical role in the
formation and acceleration of CMEs (Patsourakos \& Vourlidas 2012;
Cheng et al. 2013).

Magnetic flux ropes can emerge directly from below the photosphere
into the upper atmosphere. Using continuous vector magnetograms from
the \emph{Hinode} satellite, Okamoto et al. (2008) found that two
abutting regions with opposite polarities connected by strong
horizontal magnetic fields first grew and then narrowed, and the
orientations of the horizontal fields along the polarity inversion
line changed from the normal polarity configuration to the inverse
polarity one gradually. Moreover, there were significant blueshifts
at the strong horizontal magnetic field area. They suggested that
they observed a magnetic flux rope that was emerging from the
sub-photosphere. Helical flux ropes can also be formed through
magnetic reconnection between two bundles of J-shaped loops which
have been frequently observed as sigmoidal structures in the extreme
ultraviolet (EUV) and X-ray lines (e.g., Canfield et al. 1999;
McKenzie \& Canfield 2008; Liu et al. 2010; Green et al. 2011). In
simulations, magnetic reconnection between sheared loops are
performed due to the imposed boundary movements, and thus can form
magnetic flux ropes (Amari et al. 2000, 2003, 2011; Fan \& Gibson
2003, 2004; Aulanier et al. 2010). Moreover, using nonlinear
force-free field models, magnetic flux ropes can be reconstructed
from vector magnetic field observations (Canou et al. 2009; Canou \&
Amari 2010; Guo et al. 2010, 2013; Jing et al. 2010; Su et al. 2011;
Jiang et al. 2013; Inoue et al. 2013). After the launch of the
\emph{Solar Dynamics Observatory} ({\it SDO}; Pesnell et al. 2012),
with the help of high-quality multi-wavelength data of the
Atmospheric Imaging Assembly (AIA; Lemen et al. 2012), many authors
have reported the existence of flux ropes in the observations (Cheng
et al. 2012; Zhang et al. 2012a; Li \& Zhang 2013a, 2013b, 2013c;
Patsourakos et al. 2013).

According to some previous observational studies, flux ropes are hot
channels in the inner corona before and during solar eruptions
(e.g., Zhang et al. 2012a; Cheng et al. 2012). They can be observed
in high temperature lines (e.g., 131 {\AA}), while invisible in low
temperature lines (e.g., 171 {\AA}). Using differential emission
measure (DEM) analysis, Cheng et al. (2012) found that the
temperature of twisted and writhed flux rope is higher than 8 MK.
However, in the studies of Li \& Zhang (2013a, 2013b) and
Patsourakos et al. (2013), the flux ropes can be observed in all the
seven EUV lines formed from 0.05 MK to 11 MK. Especially, for the
two events investigated by Li \& Zhang (2013b), the flux ropes were
tracked by erupting material, leading to the visibility of them,
while they could not be detected in all wavelengths at the
pre-eruption stage.

The New Vacuum Solar Telescope (NVST; Liu \& Xu 2011) is the most
important facility of the \emph{Fuxian Solar Observatory} in China.
The diameter of NVST is 1 m and the pure aperture is 980 millimeter.
One main goal of NVST is to image the Sun at high resolution. As one
of the three channels (H$\alpha$, TiO, and G band) being used now,
H$\alpha$ is used to observe magnetic structures in the
chromosphere. In this Letter, we investigate in detail a flux rope
tracked by a filament activation using NVST H$\alpha$ observations
for the first time. Combined with the Helioseismic and Magnetic
Imager (HMI; Scherrer et al. 2012) and AIA data, we also study the
activation and the movement of the filament, and present the
corresponding appearance of higher layers revealed in different EUV
passbands.

\section{OBSERVATIONS AND DATA ANALYSIS}

The data used in this study were obtained by the NVST in H$\alpha$
6162.8 {\AA} from 02:20:02 UT to 03:20:22 UT on 2013 February 1. The
H$\alpha$ images are centered at N11$\degr$E37$\degr$ with a
field-of-view (FOV) of 159$''$ $\times$ 159$''$. The cadence of
H$\alpha$ observations is 12 s and the pixel size is 0$\arcsec$.162.
In addition, we adopt the \emph{SDO}$/$AIA multi-wavelength images
observed from 02:00 UT to 03:30 UT and \emph{SDO}$/$HMI
line-of-sight (LOS) magnetograms from 00:00 UT to 03:30 UT. Among
the 10 wavelengths of AIA, we choose 304 {\AA}, 171 {\AA}, and 131
{\AA} data with a pixel size of 0$\arcsec$.6 and a cadence of 12 s.
The HMI magnetograms have a spatial sampling of 0$\arcsec$.5
pixel$^{-1}$ and a cadence of 45 s. In addition, an NVST TiO image
and an HMI intensitygram observed at 03:10 UT are adopted for the
coalignment of observations from different instruments.

The H$\alpha$ data are dark current subtracted and flat field
corrected to Level 1, and then reconstructed to Level 1$+$ with the
speckle masking method of Weigelt (1977). All the H$\alpha$ images
are coaligned with each other by applying the image
cross-correlation method. The AIA images and HMI magnetograms are
aligned using the standard routine \emph{aia\_prep.pro}, and then
derotated differentially to a reference time (02:20 UT, 2013
February 1). The HMI intensitygram and NVST H$\alpha$ image at 03:10
UT are coaligned with the TiO image with the cross-correlation of
obvious features. Then all the AIA and HMI data are coaligned with
H$\alpha$ images.

\section{RESULTS}

Figure 1 shows the overview of the whole FOV H$\alpha$ image (left
panel) and the simultaneous HMI magnetogram (right panel). The
active region (AR) covered by the H$\alpha$ observations is AR 11665
at N11$\degr$E37$\degr$. The filament was located at the southeast
of the main sunspot. The activation of the filament began at the
northeast endpoint (outlined by square ``1") and then the filament
material moved toward the southwest direction, tracking out the
pre-existing flux rope.

\subsection{Activation of the filament}

At the area where the northeast end of the filament was located, the
magnetic fields are shown in the top panels of Figure 2. The
magnetic patches of positive and negative polarities moved toward
each other (panel (a1)) and canceled gradually (panel (a2)). Using
the inductive local correlation tracking method (ILCT; Welsch et al.
2004), we calculate the horizontal velocities of the photospheric
magnetic fields. In panel (a2),the horizontal velocities of the
negative and positive polarities represented by the red and blue
arrows indicate that the negative fields were encountering the
positive ones, resulting in the cancellation between them. At
02:28:55 UT,the magnetic fields, especially the negative polarity,
had significantly disappeared due to the cancellation (see panel
(a3)). The temporal evolution of the negative magnetic flux in the
cancellation area is shown with the read curve in panel (a3). We can
see that, from 00:13:55 UT to 02:28:55, the unsigned negative flux
persistently decreased by 26\% from 3.5 $\times$ 10$^{19}$ Mx to 2.6
$\times$ 10$^{19}$ Mx. During the cancellation process, the filament
was activated, exhibiting as rising and expanding (panels (b1) --
(b3)). At 02:29:19 UT, the fine structures of the filament can be
obviously identified and delineated with dashed curves in panel
(b3). The crossed threads indicate that the filament was twisted.
The bottom panels are the simultaneous AIA 304 {\AA} images
corresponding to the H$\alpha$ observations. At 02:20:19 UT, there
was a dark feature (denoted by the black arrow in panel (c1)) at the
filament location. While only about 7 min later, a jet-like
brightening appeared (denoted by the black arrow in panel (c2)).
Then the brightening feature expanded significantly in size, as
shown in panel (c3). The jet-like brightening is commonly considered
as a signature of magnetic reconnection.

\subsection{Filament material flow tracking out the twisted flux rope}

After activation, filament material flowed to the southwest
direction and tracked out pre-existing helical threads (see the left
column of Figure 3 and Movie 1). At 02:35:22 UT, the filament
material was mainly located near the northeast end (denoted by the
arrow in panel (a)). About 7 min later, the material moved to a
farther position (denoted by the arrow in panel (b)). At that
moment, the twisted structure outlined by the dark material was more
pronounced compared with that 7 min ago. This part of flux rope is
outlined by the blue and red curves in panel (b). The material went
on flowing and reached the place denoted by the arrow (see panel
(c)) at 02:55:57 UT. Another section of the flux rope was
consequently tracked out by the filament material, and outlined by
the blue and red curves. In order to show the entire flux rope whose
different sections were filled with dark material at different
moments, a composite H$\alpha$ image is presented in panel (d). In
panel (d), the sub-regions outlined by two green quadrangles were
observed at 02:39:12 UT and 02:42:50 UT, respectively. The rest
background image was observed at 02:55:57 UT. The entire flux rope
exhibits as a twisted structure figured out by the dark filament
material and connects the positive fields at the northeast and the
negative fields at the southwest (the red and the blue contours in
panel (d)). The approximate length of the twisted flux rope is 75
Mm. Combined with the observations at different times, the twist
configuration of the flux rope tracked out by the material flow
along the helical threads is roughly sketched out and overlaid on
the magnetogram obtained at 02:55:55 UT (see the blue and red thick
curves in panel (e)). The left part of the flux rope was tracked out
at 02:42:50 UT, and the right part was tracked out at 02:55:57 UT.
Moreover, the right section of the flux rope can be well identified
in AIA 304 {\AA} line (see the blue and green curves in Figure 4
(b2)). We estimate that the twist of the flux rope is about 1$\pi$.
To show the material flow clearly, a space-time plot along curve
``A--B" marked in panel (b) is shown in panel (f). The plot reveals
that the filament material moved from ``A" to ``B" and the mean
velocity is 31.1 km s$^{-1}$.

\subsection{Appearance in multi-wavelength images}

We also examine the multi-wavelength images obtained by AIA and
display 304 {\AA}, 171 {\AA}, and 131 {\AA} images before and during
the filament activation in Figure 4 (see also Movies 2, 3, and 4).
At 02:20 UT, before the activation, the filament appeared as a thin
dark structure in the H$\alpha$ image (panel (a1)). In 304 {\AA}
image (panel (b1)), there was a faint dark channel corresponding to
the H$\alpha$ filament, while there was no distinct dark or bright
structure in 171 {\AA} (panel (c1)) and 131 {\AA} (panel (d1))
images. At 02:46 UT, during the activation, the H$\alpha$ filament
expanded and contained many dark threads. In 304 {\AA}, the flux
rope displays as a partial bright and partial dark structure. While
in 171 {\AA} and 131 {\AA} images, there were bright structures at
the H$\alpha$ filament location.

The cuts of multi-wavelengths along slice ``A--B" marked in Figure 4
are presented in the higher panels of Figure 5. The gray shadows
mark the filament width identified in H$\alpha$ images. Panel (a)
shows the cuts at 02:20 UT. In 304 {\AA} cut, the section
corresponding to filament, i.e., the gray section in H$\alpha$ cut,
only shows a slight lower emission. While in 171 {\AA} and 131 {\AA}
cuts, there is no distinct high or low emission at the corresponding
section. Panel (b) shows the cuts at 02:46 UT. The H$\alpha$ cut
shows that there are many dark structures (see the gray section).
The cuts of 304 {\AA}, 171 {\AA}, and 131 {\AA} show much higher
emissions. Panel (b) shows a striking anti-correlation between
H$\alpha$ and EUV lines, i.e., peaks of the former correspond to
maxima of the latter and vice versa. As revealed by the H$\alpha$
cut (black curve in panel (b)), four threads of the flux rope are
well-resolved. In order to measure their width, we fit the sections
of the four threads using Gaussian fitting. The widths of them are
1.19 Mm, 0.62 Mm, 1.63 Mm, and 1.00 Mm, respectively. Their average
width of its individual threads is about 1.11 Mm.

As shown in panels (c1) and (c2) of Figure 4, there are many bright
coronal loops overlying the filament location. To examine the
influence of the filament activation on the coronal loops, a
space-time plot along curve ``C--D" (marked in Figure 4) is
displayed in panel (c) of Figure 5. We can see that, from 02:00 UT
to 03:30 UT, the coronal loops (bright structures) were quite
steady.

\section{CONCLUSIONS AND DISCUSSION}

Based on the NVST H$\alpha$ observations for the first time, we
study in detail a flux rope which was tracked by a filament
activation. The filament material was initially located at one end
of the flux rope, and the filament was activated due to the magnetic
field cancellation. The activated filament then rose and flowed
along the flux rope, tracking out the twisted structure. The length
of the flux rope is about 75 Mm, the average width of its individual
threads is 1.11 Mm, and the estimated twist is 1$\pi$. The tracked
flux rope appeared as a dark structure in H$\alpha$ images, a
partial dark and partial bright structure in 304 {\AA}, while as
bright structures in 171 {\AA} and 131 {\AA} images. During this
process, the overlying coronal loops were quite steady since the
filament was confined within the flux rope and did not erupt
successfully.

The bright core of a CME is thought to be cool filament matter which
is suspended in the dips of magnetic flux (e.g., Xia et al. 2012;
Zhang et al. 2012b). As revealed by the event in the present study,
the filament is closely related with the flux rope. It seems that,
the steady filament is only located at one section of the flux rope.
However, instead of being suspended in the dips of twisted magnetic
flux, the filament material is confined within the helical threads.
Only when the filament is activated, the filament material then can
flow easily along the threads and thus track out the twisted
structure of the flux rope. We tend to support a picture of a
pre-existing flux rope that was tracked out by filament material
activated by magnetic flux cancellation. However, since the
observation duration of the present event is very short and the
field-of-view is not large enough, we cannot exclude the formation
of new flux rope due to flux cancellation, which is a popular
mechanism for flux rope formation (e.g., van Ballegooijen \& Martens
1989).

As studied by Zhang et al. (2012a) and Cheng et al. (2012), flux
ropes could be clearly observed before and during eruptions and were
only detected in hot temperature passbands. While in this study, the
flux rope was only observed during the filament activation instead
of before the activation, which is similar to the study of Li \&
Zhang (2013b). The flux rope was tracked out by the filament
material and detected in low temperature (e.g., 304 {\AA}) and high
temperature (e.g., 131 {\AA}) lines, consistent with the results of
Li \& Zhang (2013a, 2013b). Moreover, our results show that there
exists a striking anti-correlation between H$\alpha$ and EUV lines
(see Figure 5(b)). This could imply some mild heating of cool
filament material into coronal temperatures during the filament
activation (e.g., Landi et al. 2010). The heating should not be
flare-like, reaching temperatures of 10 MK or more. This can be
demonstrated by the almost identical cuts in the AIA 171 {\AA} and
131 {\AA} channels (the 131 {\AA} channel bandpass besides
containing a flare peak at around 10 MK, also contains a ``warm"
peak at several 0.1 MK, similar to the main peak of the 171 {\AA}
channel).

In some former studies (Li \& Zhang 2013a, 2013b) and also this
study, flux ropes were observed in different EUV lines and appeared
as bright structures (see Figures 4(c2), and 4(d2)) or partial
bright structures (see Figure 4(b2)), indicating the co-existence of
hot and cool components in flux ropes. In contrary, the twisted flux
rope in H$\alpha$ images consists of dark threads, which is
different from the appearance in EUV images. For the flux ropes
studied by Li \& Zhang (2013b), the approximate length is 570 Mm.
They measured the width of individual thread of flux ropes and found
that the average width is 1.16 Mm. However, the length of flux rope
in the present study is only 75 Mm, much smaller than those studied
by Li \& Zhang (2013b). In the H$\alpha$ images, the fine-scale
threads can be generally resolved, and their mean width is
determined to be 1.11 Mm, consistent with the result of Li \& Zhang
(2013b).

\acknowledgments {This work is supported by the Outstanding Young
Scientist Project 11025315, the National Basic Research Program of
China under grant 2011CB811403, the CAS Project KJCX2-EW-T07, the
Strategic Priority Research Program$-$The Emergence of Cosmological
Structures of the Chinese Academy of Sciences (Grant No.
XDB09000000), and the National Natural Science Foundations of China
(11203037, 11221063, 11373004, and 11303049). The data are used
courtesy of NVST and SDO science teams.}

{}

\clearpage

\begin{figure}
\centering
\includegraphics
[bb=50 297 544 544,clip,angle=0,scale=0.95]
{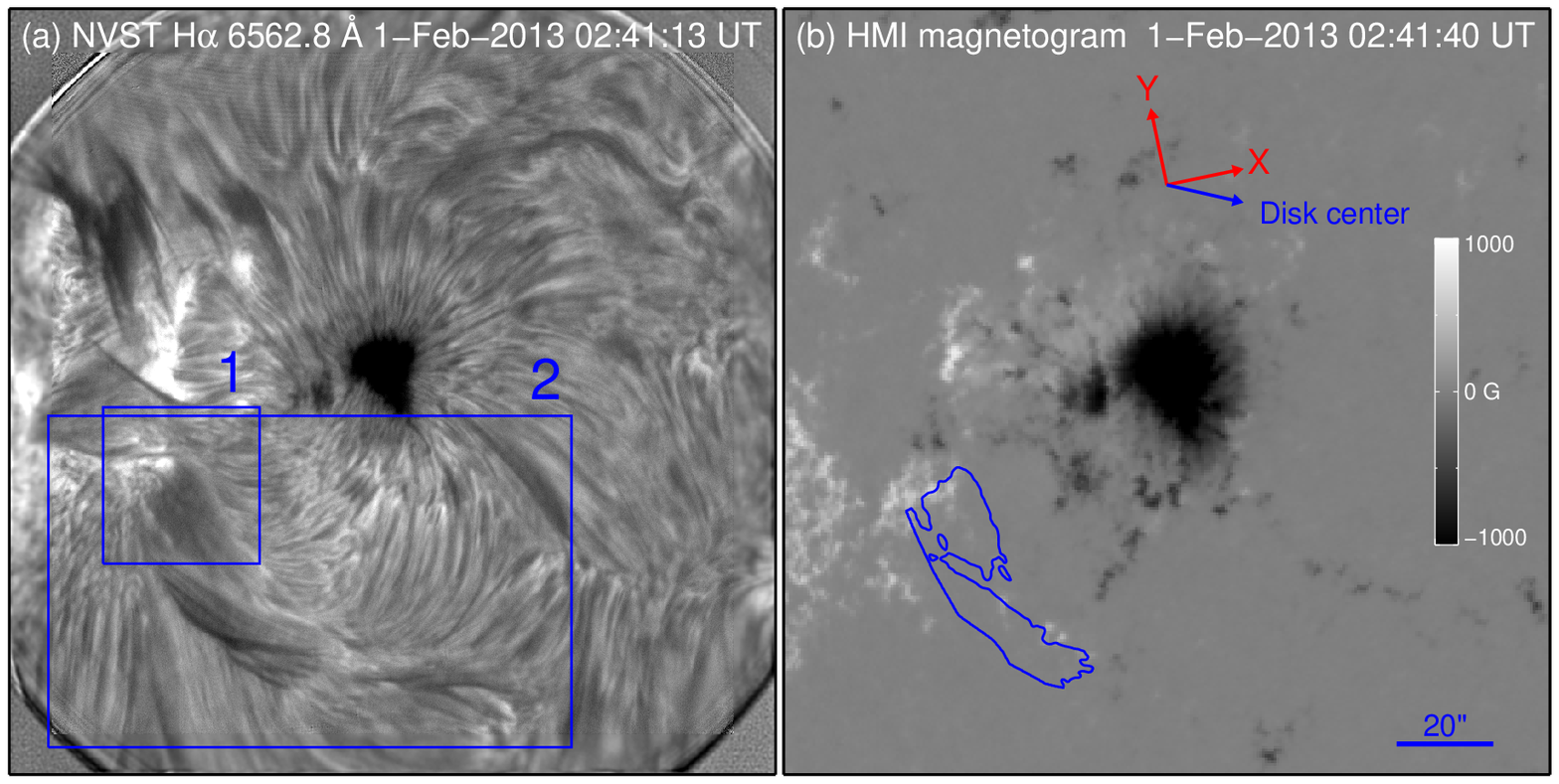} \caption{NVST H$\alpha$ image
(left panel) and HMI line-of-sight magnetogram (right panel). Square
``1" outlines the FOV of Figures 2 (b1)--(c3), and rectangle ``2"
delineates the FOV of Figures 3 and 4. The blue curve in panel (b)
is the contour of the filament during activation. \label{fig1}}
\end{figure}
\clearpage

\begin{figure}
\centering
\includegraphics
[bb=70 193 524 647,clip,angle=0,scale=1] {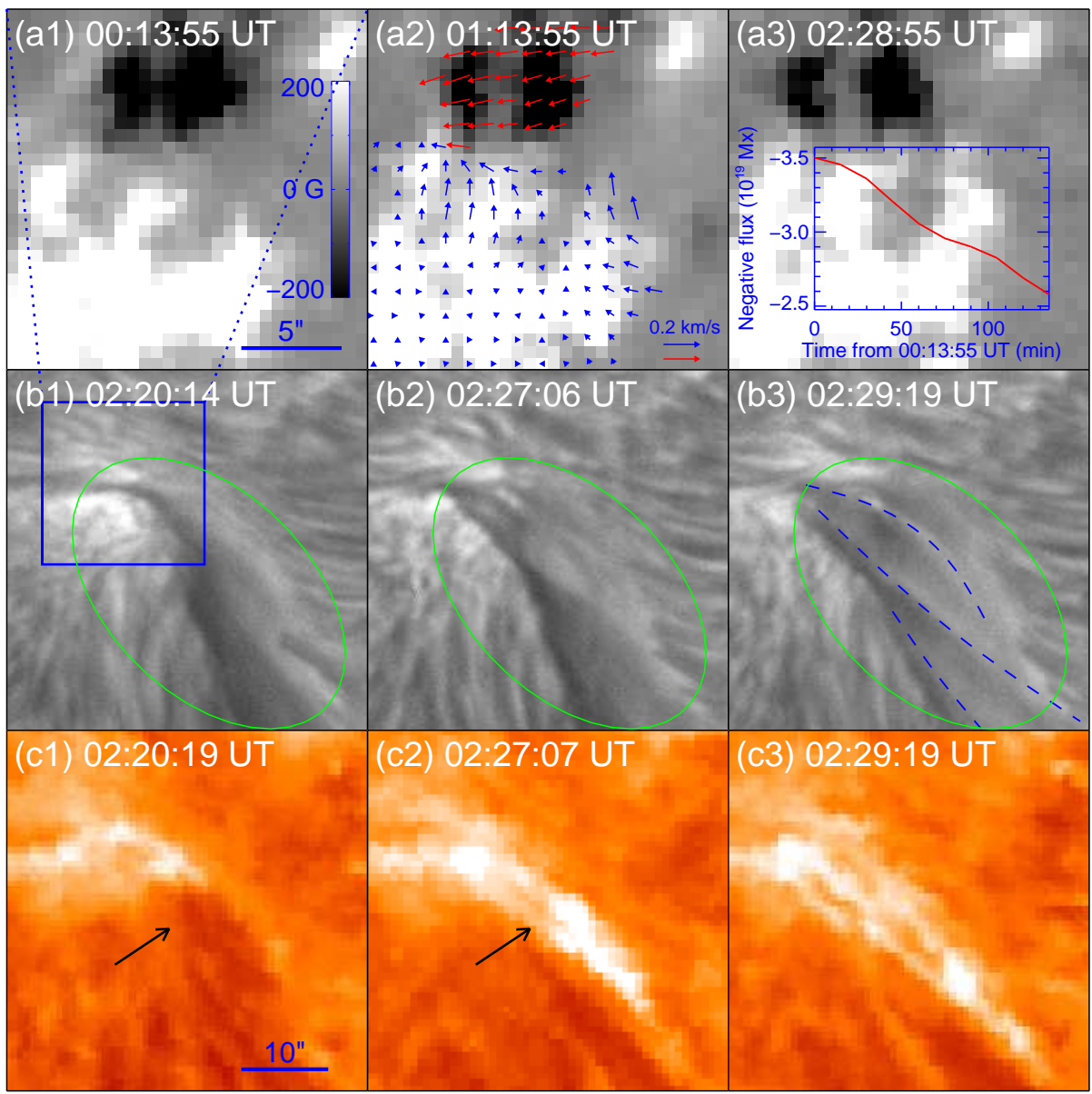} \caption{Top
panels: magnetograms showing the cancellation of opposite
polarities. Middle panels: H$\alpha$ images displaying the
activation of the filament. Bottom panels: AIA images exhibiting
brightening in 304 {\AA}. The red and blue arrows in panel (a2)
represent the horizontal velocities of negative and positive
magnetic fields, respectively. The red curve in panel (a3) displays
the temporal evolution of the negative magnetic flux in the area
where cancellation takes place. The square in panel (b1) outlines
the FOV of panels (a1--a3). The ellipses in panels (b1--b3) outline
the filament location, and the arrows in panels (c1--c2) denote the
EUV brightening. The dashed curves in panel (b3) delineate three
threads of the activated filament. \label{}}
\end{figure}
\clearpage

\begin{figure}
\centering
\includegraphics
[bb=69 188 524 636,clip,angle=0,scale=1] {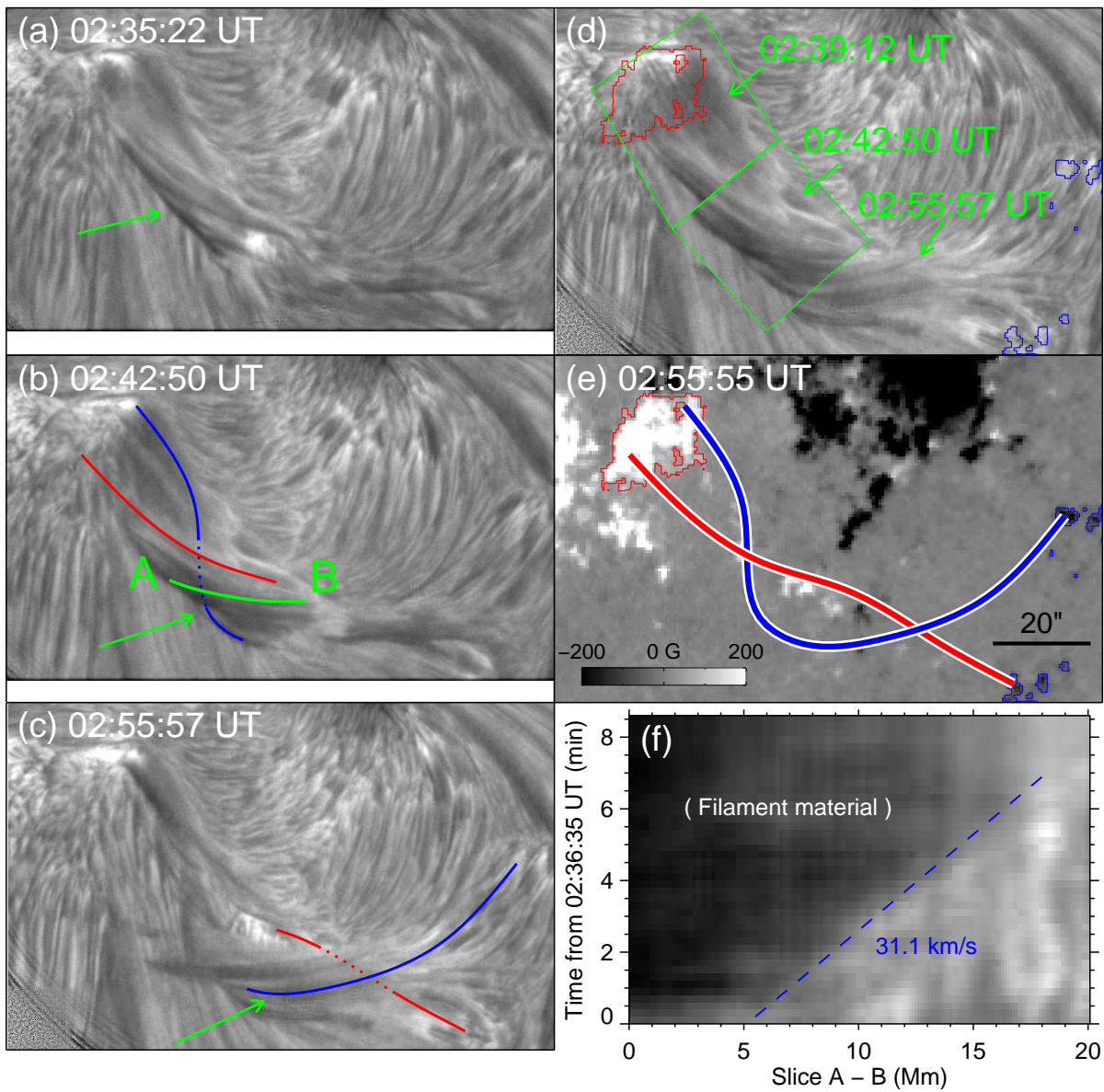} \caption{Panels
(a--c): H$\alpha$ images  showing the process of the filament
activation. Panel (d) is a composite H$\alpha$ image showing the
whole flux rope tracked by the filament material, and panel (e) is
the corresponding photospheric magnetogram. Panel (f) is the
space-time plot showing the dark filament material flow along curve
``A--B" marked in panel (b). The arrows in panels (a--c) denote the
filament locations at different stage. The red and blue thin curves
in panels (b) and (c) outline the left part and right part of the
flux rope, respectively, while the thick curves in panel (e) display
the whole twist configuration of the flux rope. The red and blue
thin curves in panels (d) and (e) indicate the flux rope footpoints
with positive and negative polarities, respectively. \label{}}
\end{figure}
\clearpage

\begin{figure}
\centering
\includegraphics
[bb=71 134 522 706,clip,angle=0,scale=1] {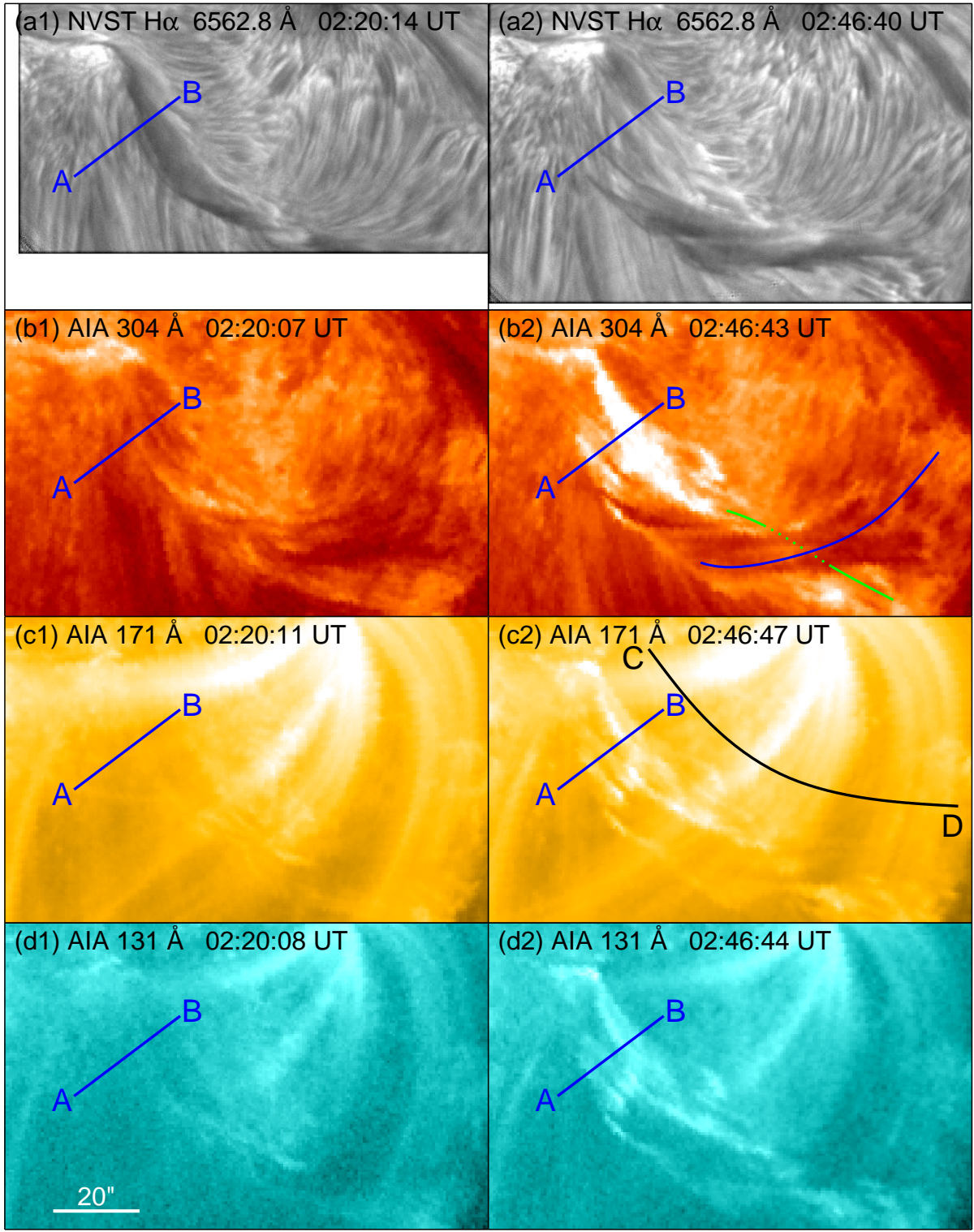}
\caption{Multi-wavelength images before (left panels) and during
(right panels) the filament activation. Lines ``A"--``B" and curve
``C"--``D" mark the positions along which the cuts and the
space-time plot in Figure 5 are obtained. The green and blue curves
in panel (b2) outline the right part of the flux rope that can be
well identified in AIA 304 {\AA} line. \label{}}
\end{figure}
\clearpage

\begin{figure}
\centering
\includegraphics
[bb=92 206 473 609,clip,angle=0,scale=1.1] {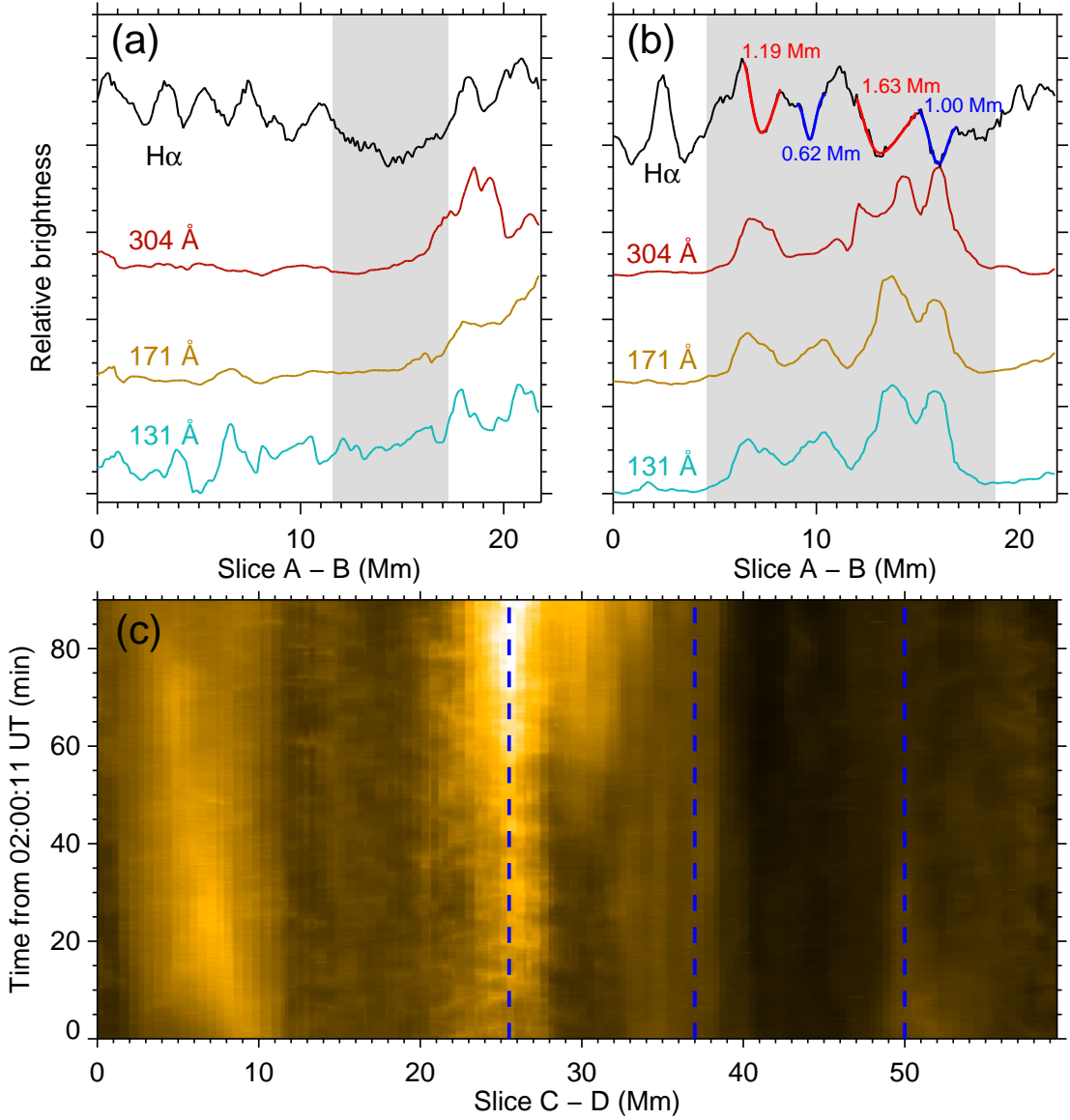}
\caption{Panels (a) and (b): relative brightness of
multi-wavelengths along slice ``A--B" (see Figure 4) at 02:20 UT and
at 02:46 UT, respectively. Panel (c): space-time plot along curve
``C--D" marked in Figure 4. The red and blue curves overlaid on the
black curve in panel (b) are Gaussian fittings of four flux rope
threads. \label{}}
\end{figure}
\clearpage

\end{document}